\documentstyle[12pt,rotating,epsfig,glas]{article}
\begin{document}
\def\AP#1#2#3{Ann.\ Phys.\ (NY) #1 (19#3) #2}
\def\CPC#1#2#3{Computer Phys. Comm. #1 (19#3) 32}
\def\PL#1#2#3{Phys.\ Lett.\ #1B (19#3) #2}
\def\PR#1#2#3{Phys.\ Rev.\ #1D (19#3) #2}
\def\NP#1#2#3{Nucl.\ Phys.\ #1B (19#3) #2}
\def\NC#1#2#3{Nuovo Cimento\ #1A (19#3) #2}
\def\ZP#1#2#3{Z.\ Phys.\ C#1 (19#3) #2}
\def\JL#1#2#3{JETP Lett.#1 (19#3) #2}
\def\JP#1#2#3{J.\ Phys.\ G#1 (19#3) #2}
\def\NIM#1#2#3{Nucl.\ Instr.\ Meth. A#1 (19#3) #2}
\def\RMP#1#2#3{Rev.\ Mod.\ Phys. #1 (19#3) #2}
\newcommand{\B}{\mbox{$B$}}

\begin{titlepage}{GLAS-PPE/97--09}{October 1997}
\title{CP Violation and
Future `B-Factories'}
\author{ N.~H.~Brook}
\conference{Lecture given at ``The Actual Problems of Particle
Physics'', Gomel, Belarus.}
\begin{abstract}
This lecture contains a brief introduction to 
CP violation in the \B\ system before discussing
future experimental programmes and their CP reach in the \B\ system.
\end{abstract}
\end{titlepage}
% \newpage
\section{CP Violation in the B-system}
Quark mixing  in the standard model is described by the
Cabibbo-Kobayashi-Maskawa (CKM) matrix~\cite{CKM}, eqn(\ref{eq:CKM}).
Conventionally the $u,c {\rm\ and\ } t$ quarks are unmixed and the 
mixing is described by the $3\times 3\; V_{CKM}$ matrix operating on the 
$d,s {\rm\ and\ } b$ quarks.
The matrix elements of $V_{CKM}$ can, in principle,
be determined by measuring the charged current
coupling to the $W^{\pm}$ bosons.
\begin{equation}
V_{CKM} \equiv \left(
\begin{array}{ccc}
V_{ud} & V_{us} & V_{ub} \\
V_{cd} & V_{cs} & V_{cb} \\
V_{td} & V_{ts} & V_{tb} 
\end{array}
\right)  
\label{eq:CKM}
\end{equation}

The CKM matrix is unitary ie $ V_{CKM}^{\dagger}V_{CKM} = 1,$ which
leads to 9 unitarity conditions expressed in terms of the matrix elements.
There are several (approximate)
parameterisations of the CKM matrix, one of the more
popular approaches is that of Wolfenstein~\cite{Wolfie},
eqn(\ref{eq:wolfie}), where the matrix elements are expressed in terms
of powers of $\lambda = \sin \theta_c,$ where $\theta_c$ is the Cabibbo angle. 
As can be seen from this parameterisation,  the
9 complex elements of the matrix can be expressed in terms of 4 independent
variables; three real parameters $A, \lambda {\rm \ and\ }\rho$ and an
imaginary part of a complex number, $\eta.$ 
The 18 parameters of the CKM matrix can be reduced to 4 because of the
unitarity constraints and the arbitrary nature of the relative quark
phases~\cite{rosner}.
It is the complex phase in the $V_{CKM}$ that leads to CP
violation in the standard model.

\begin{equation}
 V_{\rm Wolfenstein} = \left(
\begin{array}{ccc}
1-\frac{1}{2}\lambda^2 & \lambda & A\lambda^3(\rho - i\eta)\\
-\lambda & 1-\frac{1}{2}\lambda^2 & A\lambda^2 \\
A\lambda^3(1-\rho - i\eta) & -A\lambda^2 & 1
\end{array}
\right)
\label{eq:wolfie}
\end{equation}

\begin{figure}[hbt]
% \centerline{\epsfig{figure=uni_tri_new.eps,height=8cm,angle=-90.}}
\vspace*{3.0cm}
\centerline{\epsfig{figure=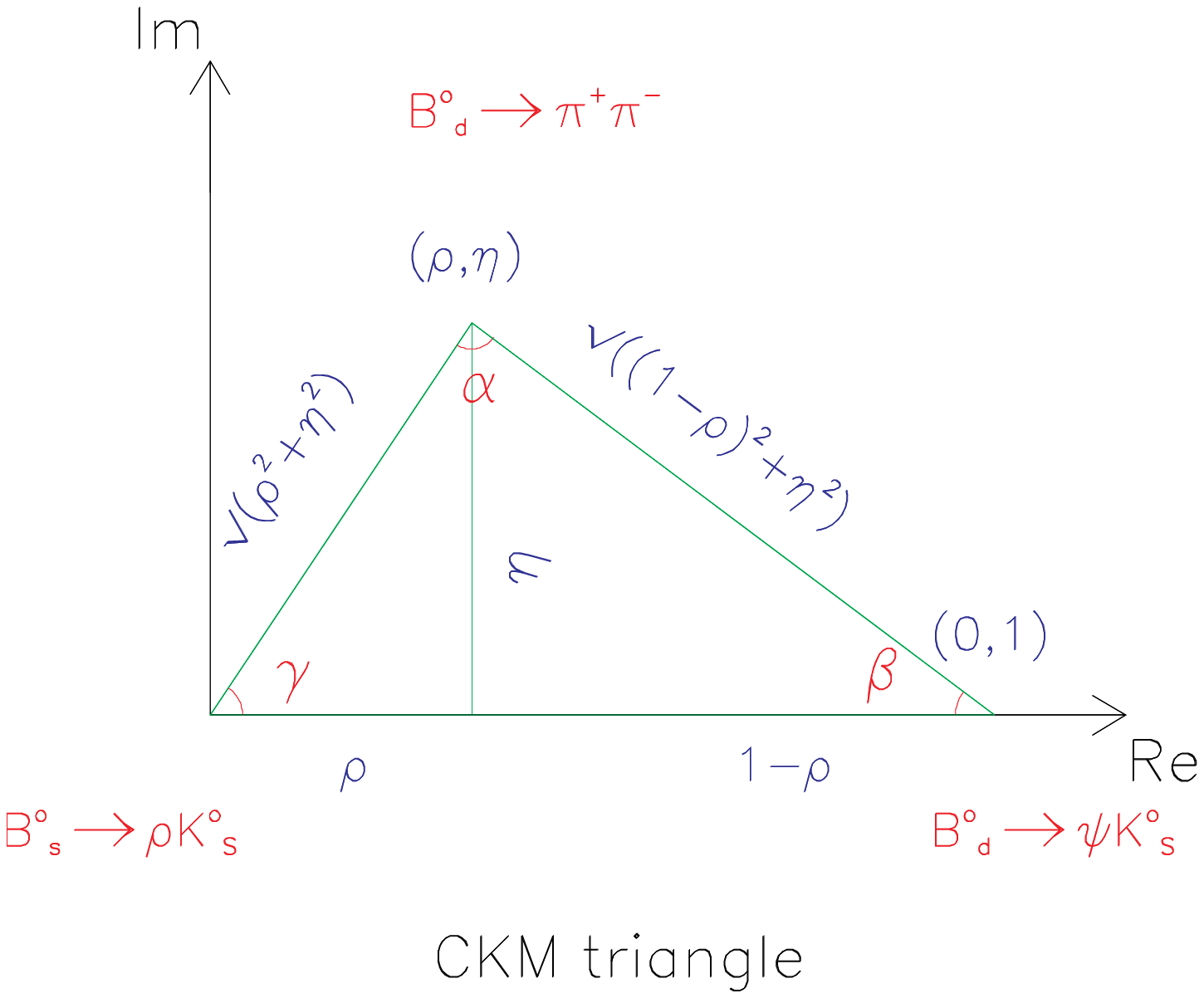,width=5cm}}
\vspace*{2.5cm}
\caption{\it The CKM unitarity triangle in the Wolfenstein parameterisation}
\label{fig:unitri}
\end{figure}

The unitarity condition 
$$V_{ud}V_{ub}^* + V_{cd}V_{cb}^* + V_{td}V_{tb}^* = 0 $$
is of particular interest since $V_{ud} \simeq V_{tb} \simeq 1$ and 
$V_{ts}^* \simeq -V_{cb}.$ This allows us to depict this condition as a
triangle in the complex plane, as shown in fig~\ref{fig:unitri}.
The angles of the triangle $\alpha, \beta {\rm \ and\ } \gamma$ are
related to the phase and can be measured in CP violating $B$-decays.

The non-closure of this
triangle ie $\alpha + \beta + \gamma \ne \pi$ would suggest that our
understanding of CP violation within the Standard Model was incomplete.
Physics beyond the Standard Model can be further investigated, for example, 
by measuring CP asymmetries in several $B$ decays that depend on the
same unitarity angle or studying decays where zero asymmetries are
expected in the Standard Model. 

\begin{figure}[htb]
\centerline{\epsfig{figure=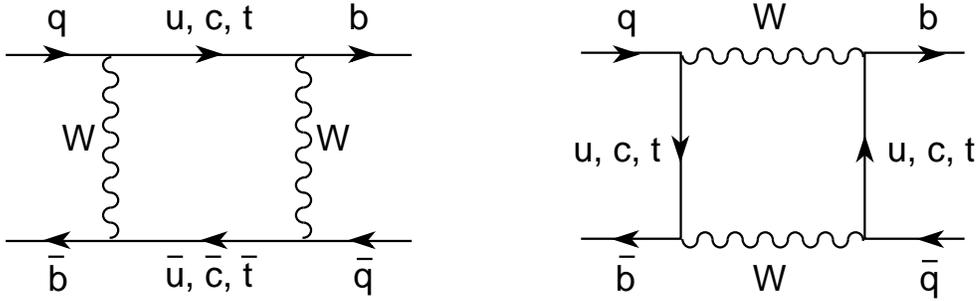,width=13cm}}
\caption{\it $B^0-\bar{B^0}$ mixing diagrams.}
\label{fig:bbar}
\end{figure}

CP violation in the \B\ system should be observable through the phenomenon
of $B^0 - \bar{B^0}$ mixing, see for example~\cite{nir_quinn}. This $B^0 -
\bar{B^0}$ mixing is dominated by box-diagrams with virtual $t-{\rm quarks},$
fig~\ref{fig:bbar}.

\newpage
The following decays

\begin{eqnarray}
B^0_d & \rightarrow & J/\psi K^0_s \nonumber \\
B^0_d & \rightarrow & \pi^+ \pi^- \label{eq:decay}\\
B^0_s & \rightarrow & \rho K^0_s \nonumber
\end{eqnarray}

\noindent are into a CP eigenstate. 
If this is coupled with only a single diagram
contributing to the decay,  CP asymmetries can be constructed which are
directly related to the angles of the unitarity triangle. For example,
these conditions occur 
for the decay mode $B^0_d \rightarrow J/\psi K^0_s.$
Here the number of $B^0_d$ which decay at time $t$ (where $t$ is
expressed in units of lifetime) is proportional to
\begin{equation}
 n(t) \propto e^{-t}(1 + \sin 2\beta \sin xt)\label{eq:numb_b}
\end{equation}
and the number of $\bar{B^0_d}$ is proportional to
\begin{equation}
 \bar{n}(t) \propto e^{-t}(1 - \sin 2\beta \sin xt)\label{eq:numb_bbar}
\end{equation}
where the mixing parameter, $x=\Delta M/\Gamma\simeq 0.67,$ 
is the ratio of the mass difference of the eigenstates to their decay rate.
The CP asymmetry can then be defined as 

% \begin{equation}
$$ a(t) = \frac{n(t) - \bar{n}(t)}{n(t) + \bar{n}(t)} = \sin 2\beta \sin xt.  $$
% \label{eq:cpasymm}
% \end{equation}

\begin{figure}[hbt]
\centerline{\epsfig{figure=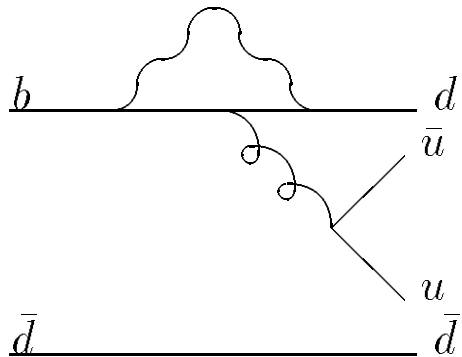,height=5cm}}
\caption{\it `Penguin' contribution to $B^0_d\rightarrow \pi^+\pi^-$ decay}
\label{fig:peng}
\end{figure}

% For \B\ meson pairs 
% not produced in a state of definite CP, 
By
integrating eqns.~(\ref{eq:numb_b}) and (\ref{eq:numb_bbar}) 
over time a similar asymmetry can be constructed which is
proportional to
$\sin 2\beta$. (Although
for coherent \B\ production ie the $B
\bar{B}$ pair is produced in a definite CP state, this time
integrated asymmetry is zero.)
In addition this channel is experimentally very promising
because of the dilepton decay of the $J/\psi.$
Unfortunately  additional decay diagrams contribute to the other
channels listed in eqn(\ref{eq:decay})
so there is no longer a complete cancellation of
the hadronic matrix elements in the CP asymmetry. The $B^0_d
\rightarrow \pi^+ \pi^-$ channel, which is dependent on the angle
$\alpha$ is predicted to have large hadronic corrections from
`penguin' diagrams, fig.~\ref{fig:peng}. The $B^0_s
\rightarrow \rho K^0_s$ (dependent on the angle $\gamma$) also has
additional hadronic contributions but, in addition, suffers from
a very low branching fraction.  Fortunately 
there are, of course, many other channels which can be used to measure
CP violation, eg see ref.~\cite{dunietz}.

\section{B-Production Facilities}
Because of the small visible branching ratios of decays to CP
eigenstates, ${\cal O}(10^{-5}),$ a large
number of \B\ meson must be produced in order to study CP violation in the
\B\ system. There are two complimentary ways to achieve the necessary
large number of
\B\ mesons. The first is at $e^+e^-$ colliders at centre of mass energy
of 10 GeV
to produce the $\Upsilon(4S)$ which then decays to two \B\ mesons. 
Alternatively 
high energy hadron machines, where there is a large cross section for
$b\bar b$ production, can be used to  produce
the \B\ mesons. The pros and cons of the various approaches are discussed
in this sections.

\subsection{$\boldmath e^+e^-$ Colliders}
\B\ meson production via $e^+e^-$ colliders
is being pursued at laboratories in
the United States (Cornell and SLAC) and Japan (KEK). The luminosity
at these machines is of the ${\cal O}(10^{33}){\rm \
cm^{-2}s^{-1}}$ which is equivalent to approximately 4 $b\bar b$ pairs
produced every second. 
Because the \B\ mesons produced from the decay of
$\Upsilon(4S)$ are coherent it is necessary to  be able to measure the
time separation between the two B's 
in order to measure the CP asymmetry through $B^0 - \bar{B^0}$ mixing.
To give the $\Upsilon(4S)$ sufficient boost to allow the two
$B^0$ decay vertices to be reconstructed and thus the distance between
the two $B$ mesons to be measured,
 the beam
energies at SLAC and KEK are asymmetric. 
The Cornell B-facility has
symmetric beam energies and will be unable to measure CP asymmetry
through $B^0 - \bar{B^0}$ mixing though there are possibilities to
measure CP violation through the decays of charged \B's. The KEK and
SLAC facilities are asymmetric with $e^-(e^+)$ energies of 3.5~GeV (8 GeV)
and 3.1~GeV (9 GeV) respectively. The need for the large luminosity and
the asymmetric beam energies poses great challenges on the machine
design. The advantages of this approach is the very clean production
environment of the 2 \B\ mesons, with no underlying event from which
to extract the signal.
By running at the mass of $\Upsilon(4S)$ it is not possible, simply 
by kinematic contraints, to
 study the $B_s$ system. In order to study the $B_s$ system the machine
can be
operated with energies at the mass of $\Upsilon(5S)$ but the
cross sections are much smaller.

\subsection{Hadron Colliders}
The high energy hadron machines, Tevatron and HERA (in fixed target mode
with a wire target inserted into the proton beam halo), are already
producing large numbers of B mesons. The LHC will produce even greater
numbers. 
At the Tevatron the $b \bar b$ production rate is
${\cal O}(10^4){\rm\ Hz}$ and at LHC it will be ${\cal O}(10^5){\rm\ Hz}$
compared to the $\simeq 4{\rm\ Hz}$ at SLAC and KEK.
The problem in this approach is achieving high enough
reconstruction and tagging efficiencies in order to extract a sufficient
number of \B's to measure CP violation. The LHC has an additional
benefit over the
Tevatron and HERA because with the increasing  centre of mass energy the
ratio of the $b \bar b$ cross section over the total inelastic cross
section also increases. 
All the hadron machines also have the advantage that they can study
$B_s$ mesons.

\section{The Experiments}

\begin{figure}[hbt]
\centerline{\epsfig{figure=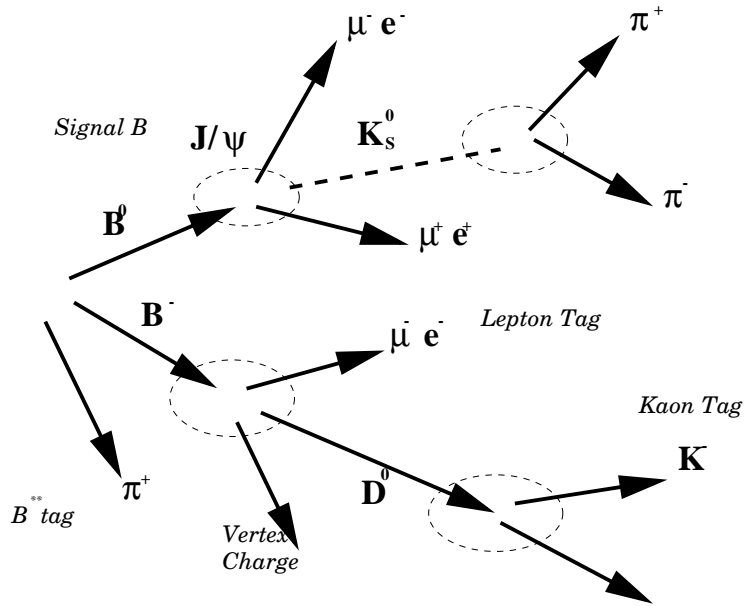,width=8cm,angle=-90}}
\caption{\it Example of B tagging.}
\label{fig:btag}
\end{figure}

To measure CP violation it is not only necessary to measure the
decay of the \B\ meson but also to tag its initial flavour via 
the decay of the
accompanying \B, as shown schematically in fig.~\ref{fig:btag}.
The generic detector requirements for an experiment to study \B\ decays are
that it has good resolution on measuring the decay time and the mass of
decayed \B\ mesons and has particle identification to allow a good initial
flavour tag of the \B\ meson. The characteristics of the detectors
designed to study CP violation are discussed below.

\subsection{BELLE}

\begin{figure}[hbt]
\centering
{\epsfig{figure=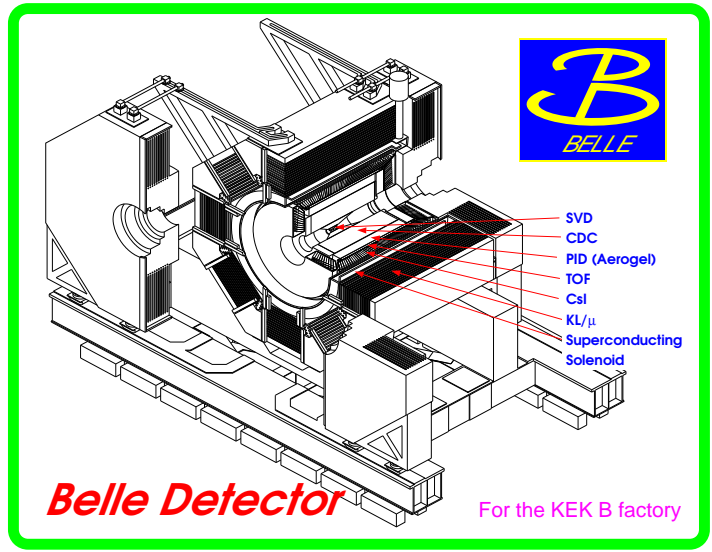,width=8.0cm,angle=-90}}\quad
\caption{\it Schematic of the BELLE detector at the KEK B factory.}
\label{fig:belle}
\end{figure}

The BELLE detector~\cite{belletdr}
 is the experimental apparatus being designed for the
KEK B factory. A schematic of the detector is shown in
figure~\ref{fig:belle}. 
The detector consist of a silicon vertex detector (SVD) situated just
outside the beampipe. Surrounding that there is a cylindrical wire drift
chamber (CDC) that measures charged tracks which extends to a radius 
of 90~cm. Particle identification is provided by $dE/dx$ measurements in
the CDC, and aerogel \v{C}erenkov counter and time of flight (TOF) arrays
situated radially around the CDC. Inside the superconducting
solenoid is a electromagnetic calorimeter manufactured from 
${\rm CsI(T}l)$ crystals. The iron return yoke of the 1.5 Tesla solenoid
is interspersed with arrays of detectors for measuring muons and $K^0_L$
mesons. The design of the equivalent detector, BaBar, at the SLAC B factory 
is discussed in ref.~\cite{babartdr}.

\subsection{HERA-B}

\begin{figure}[hbt]
\centering
{\epsfig{figure=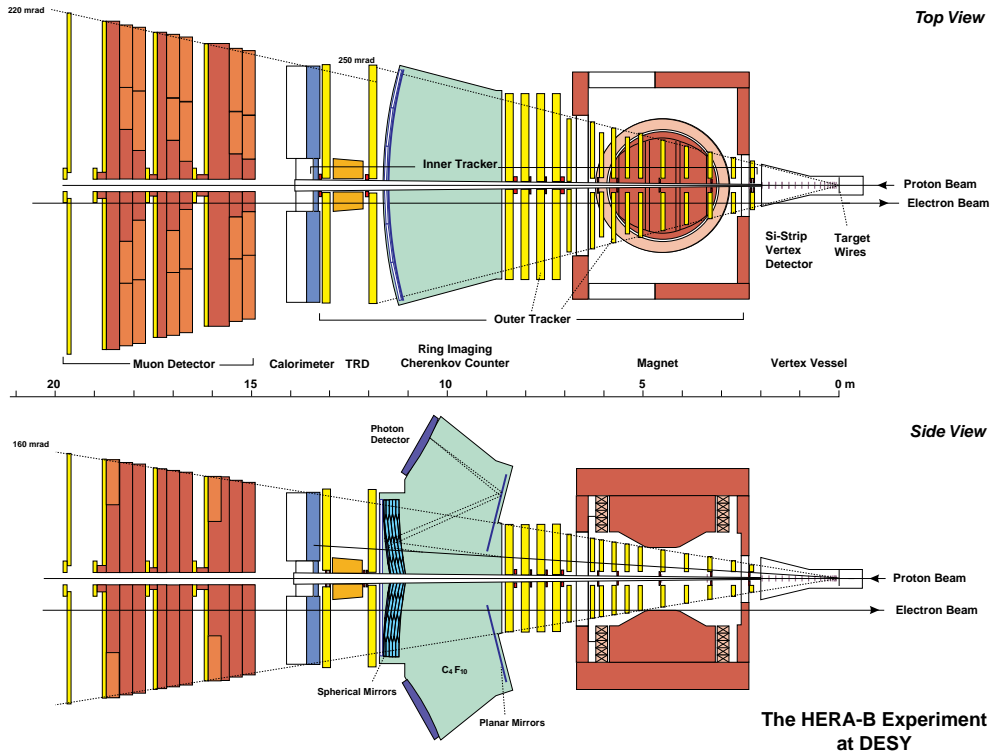,width=10.0cm,angle=-90}}
\caption{\it Schematic of the HERA-B detector at DESY.}
\label{fig:herab}
\end{figure}

To guarantee the observation of standard model CP violation
in \B\ decays (after folding in the detector efficiencies),
interactions at the HERA-B detector have to occur
approximately 4 times for every one of the  10~MHz bunch crossings of the HERA
machine. The HERA-B experiment is essentially a fixed target experiment
with a wire target in the beam halo~\cite{herabtdr}. 
A schematic of the HERA-B
spectrometer is shown in fig.~\ref{fig:herab}.
It has a single dipole momentum spectrometer
situated 4.5~m downstream of the target.
Directly downstream of the target wires, but before the magnet, there is
a silicon vertex detector
of length of $\sim 2{\rm \ m}.$ The main tracking system uses a variety
of technologies dependent on the distance away from the beam (Si-strips,
microstrip gas counters and honeycomb-drift chambers at ever increasing
radii from the beam.) This is followed by a ring imaging \v{C}erenkov
(RICH) detector to tag the charged kaon and a transition radiation detector  to
improve electron identification. There are additional large tracking
chambers immediately behind the  RICH and in front of the
calorimeter. The electromagnetic calorimetry is designed to use
Lead/Scintillator and Tungsten/Scintillator. This is followed by a
conventional muon system with four chamber layers at various depths in
the absorber. The muon chambers are essential for the triggering
of HERA-B when the $J/\psi$ decays to two muons.

\subsection{CDF and D0}
\begin{figure}[ht]\vspace{-5mm}\begin{center}
\mbox{
\epsfig{file=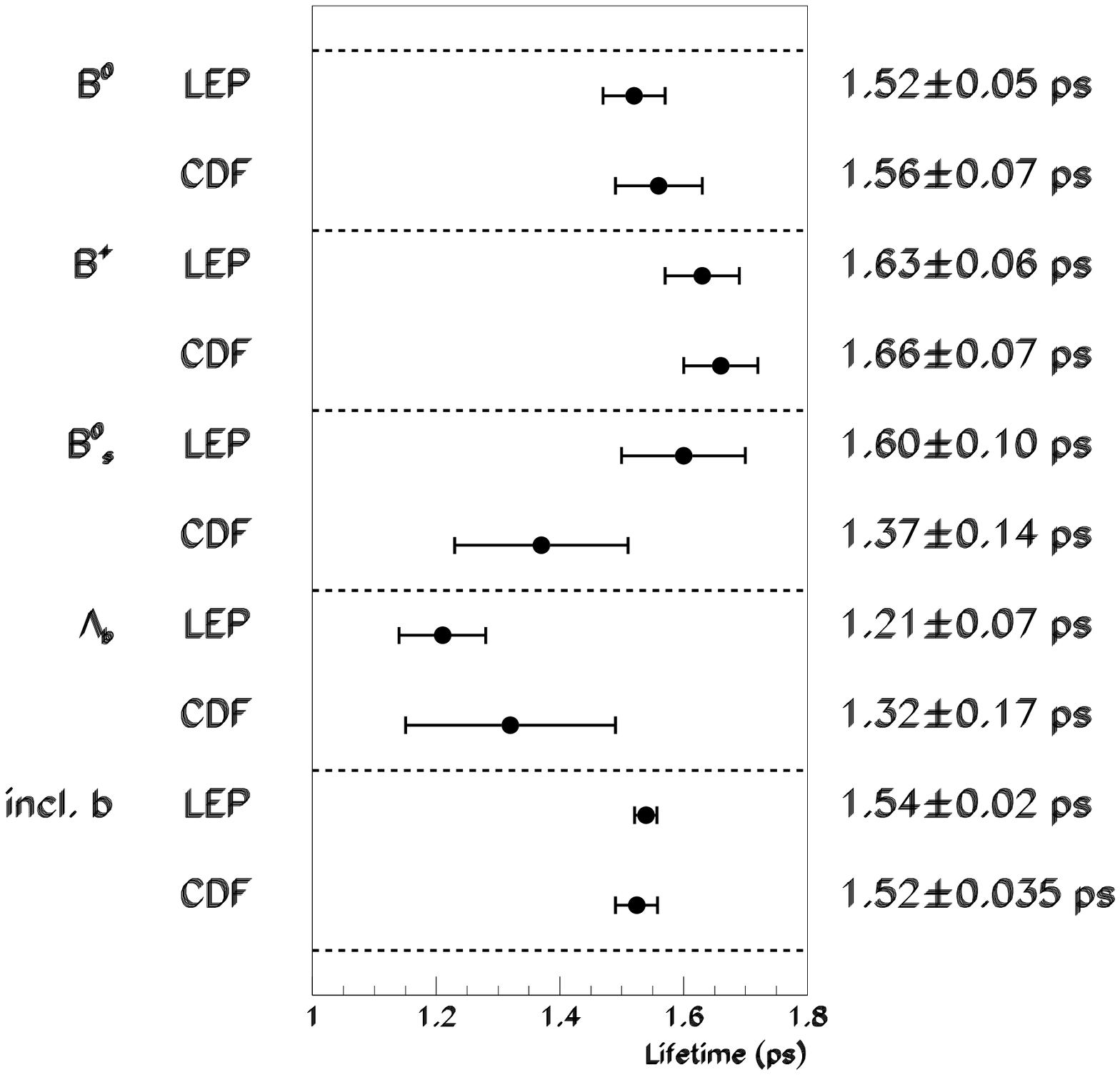,clip=yes,width=0.4\textwidth}
\hspace{-1mm}
\epsfig{file=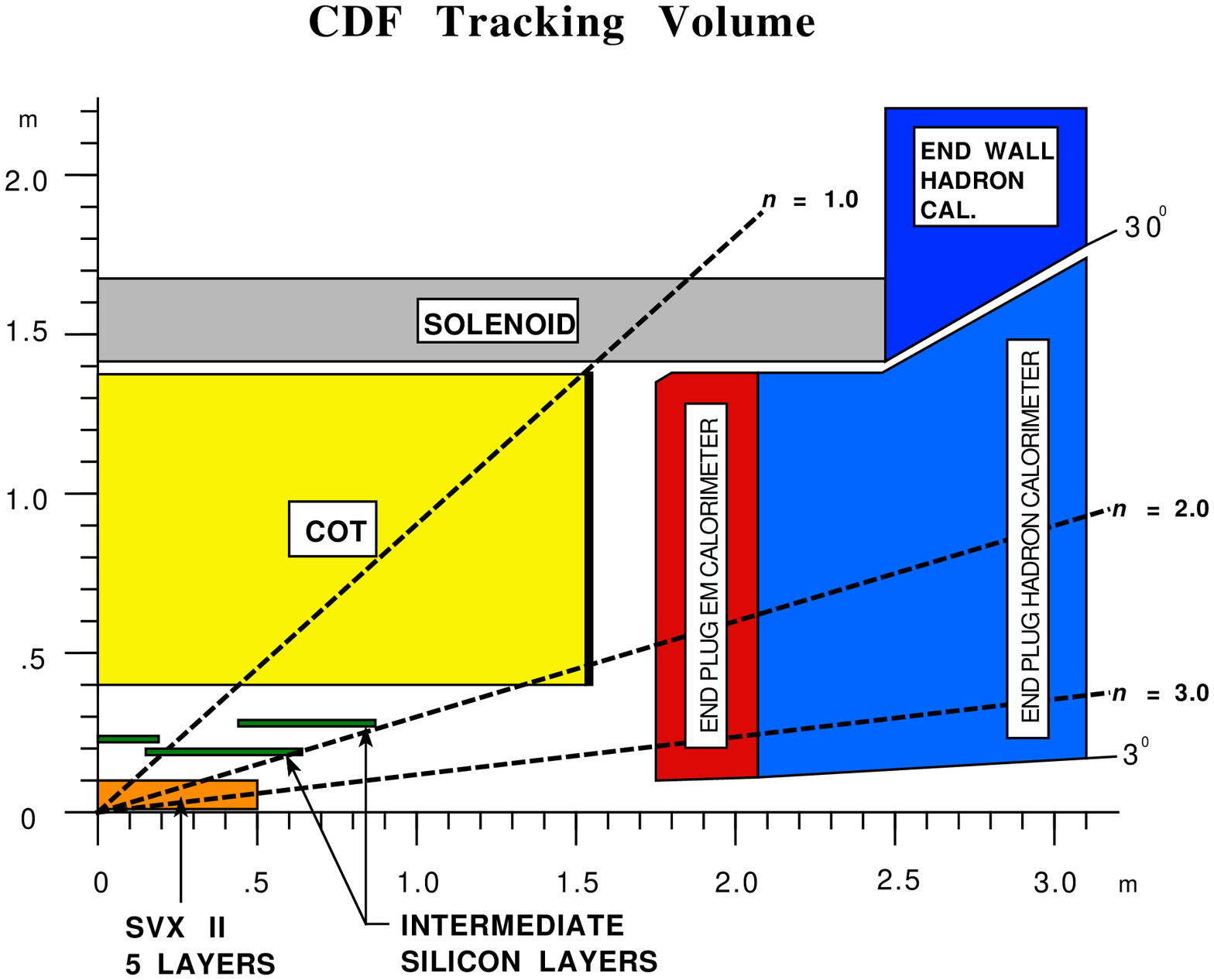,clip=yes,width=0.6\textwidth}}
\end{center}
\vspace{-1.1cm}
\begin{minipage}[t]{0.4\textwidth}
\caption{\it A comparison of CDF and LEP \B\ meson lifetime measurements.}
\label{fig:blife}
\end{minipage}\hfill
\begin{minipage}[t]{0.54\textwidth}
\caption{
\it Longitudinal View of the CDF II Tracking System.
  }\label{fig:cdfup}
\end{minipage}
\end{figure}

There is already a very active \B\ physics program at the CDF detector at the
Tevatron. The \B\ meson lifetime measurements, fig.~\ref{fig:blife}, are
already competetive with those of the combined LEP experiments. 
This proves
that it is possible to extract \B\ physics from the hostile experimental
environment of the hadron colliders. Both CDF~\cite{cdfup} and D0~\cite{d0up}
 are hoping to
exploit the Tevatron upgrade (RUN II in 1999)
to study CP violation in the \B-system.

% \begin{figure}[hbt]
% {\epsfig{figure=lifetime.eps,width=5.0cm}}
% \caption{\it A comparison of CDF and LEP B meson lifetime measurements.}
% \label{fig:blife}
% \end{figure}

CDF are proposing a new tracking system for Run II, fig.~\ref{fig:cdfup}. 
At large radii they
will have a central outer tracker (COT) of an open drift chamber design.
Inside this component there is a silicon inner tracker compromising of
two components: a micro-vertex detector (SVX II) at very small radii and two
additional layers of silicon at intermediate radii.
The current forward calorimetry is going to  be replaced with a new
scintillating tile plug calorimeter. New chambers will be added to the
current muon system to close gaps in the azimuthal acceptance and the
forward acceptance will be improved.
The trigger will also be upgraded to allow track finding at level-1
and the ability to trigger on large impact parameter tracks at level-2.

A major element of the D0 upgrade is their new inner tracking system
% (fig.~\ref{fig:cdfup})
which will be located insides a new 2 Tesla superconducting solenoid. It
will consist of an inner silicon vertex detector, surrounded by eight
superlayers of scintillating fibre tracker.
In the forward region, on the face of the end calorimeter cryostats,
and in the central region, located between the solenoid and the inner
radius of the calorimeter cryostat, a scintillator based pre-shower
detector will be installed. The trigger upgrades will include tracking
triggers at level-1 and an upgraded level-2 muon triggers.

\subsection{ATLAS and CMS}
The ATLAS~\cite{ATLASTDR} and CMS~\cite{CMSTDR}
 detectors at the LHC are optimised for high luminosity
physics. But initial low luminosity running will allow these general
purpose detectors to used for \B\ physics. Both detectors are installing
silicon vertex detectors close to the beampipe. More detailed discussion
of these detectors and their physics performance can be found elsewhere
in these proceedings~\cite{theseproc}.

\begin{figure}[hbt]
\centering
{\epsfig{figure=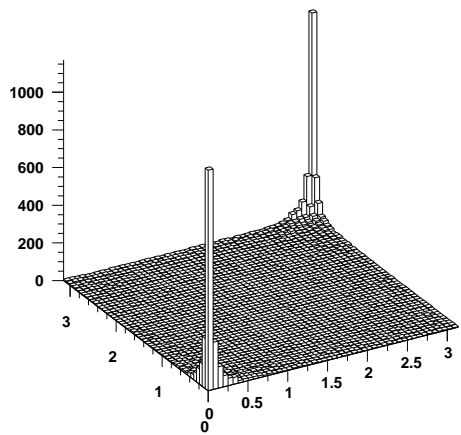,height=6.0cm,bbllx=40pt,bblly=60pt,bburx=530pt,bbury=515pt }}
\caption{\it Production angle of B vs. angle of $\overline{\rm B}$ in the
laboratory (in units of rad.) at LHC calculated using the the PYTHIA Monte
Carlo generator~\protect\cite{PYTHIA}.}
\label{fig:bang}
\end{figure}

\subsection{LHC-B}
At high energy hadron colliders
the produced $B$ and $\bar B$ mesons are correlated  in the
forward direction (ie close to the proton beam direction).
Figure~\ref{fig:bang} shows the angular distribution of the $B \bar B$
mesons in the laboratory frame at LHC. This is due to the relatively low mass
production of $b$ quark pairs at collider energies. This production
mechanism lends itself to dedicated experiments that are of a forward,
planar design, reminiscent of those used in fixed target experiments.

\begin{figure}[htb]
\centering
{\epsfig{figure=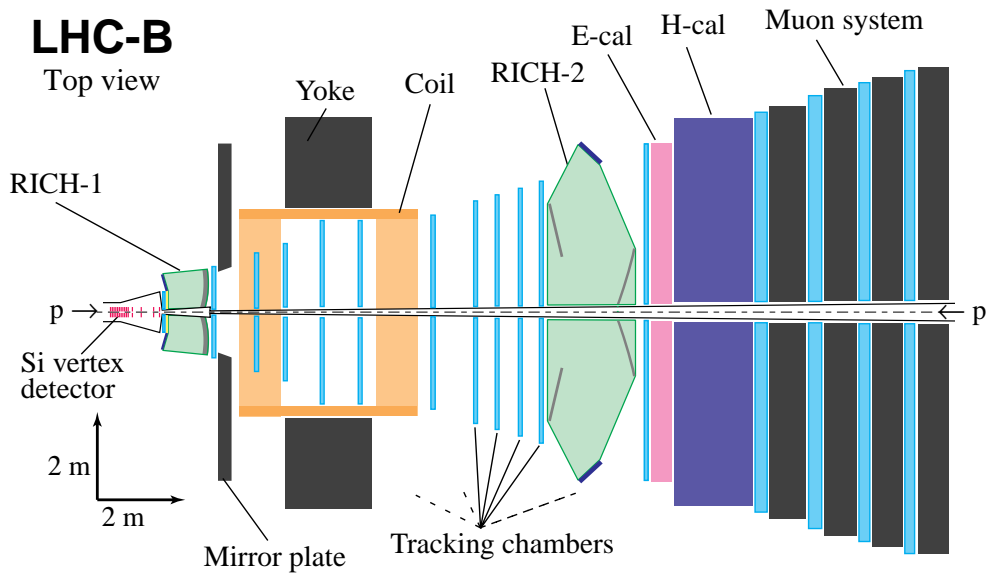,width=13.0cm }}
\caption{\it Top view of the LHC-B detector}
\label{fig:lhcb}
\end{figure}

Such an experiment, LHC-B, has been proposed for the LHC~\cite{LHCBLoI}. 
Its layout is
shown in figure~\ref{fig:lhcb}. LHC-B is a forward single-dipole
spectrometer It consists of a silicon microvertex detector, a tracking
system, aerogel and gas RICH detectors, electromagnetic and hadronic
calorimeter and a muon filter.
LHC-B will be allowed to run with a defocussed beam in their interaction
area which would give a nominal running luminosity ${\cal L} = 1.5 \times
10^{32}{\rm\ cm^{-2}s^{-1}.}$

\subsection{BTeV}
\begin{figure}[hbt]
\centering
{\epsfig{figure=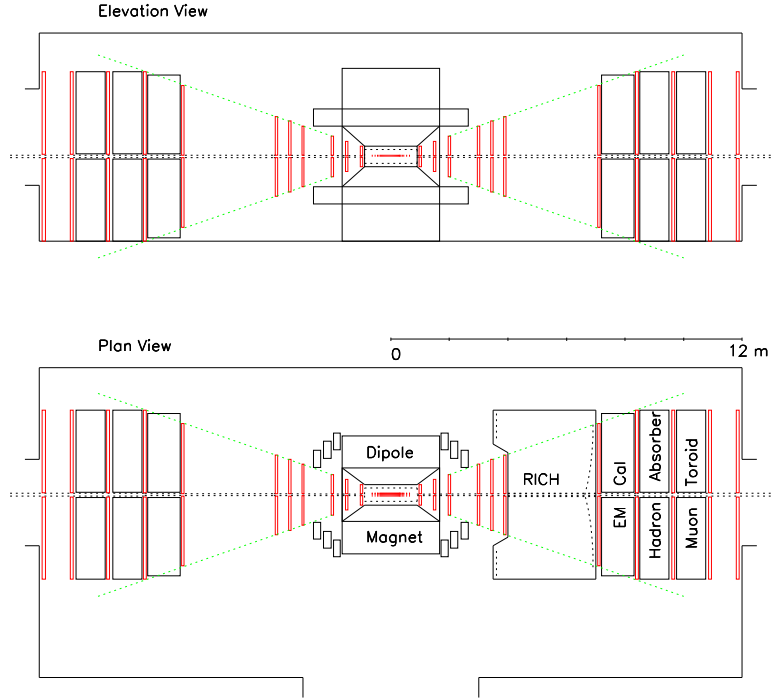,height=10.0cm,bbllx=0pt,bblly=55pt,bburx=415pt,bbury=450pt }}
\caption{\it Schematic Layout of the BTeV detector}
\label{fig:btev}
\end{figure}

Recently a similarly motivated experiment (BTeV) has submitted a expression of
interest at Fermilab~\cite{BTeVEoI}. 
The schematic layout of the BTeV proposal is
shown in figure~\ref{fig:btev}. The baseline description of the detector
has a dipole magnet centred on the interaction region, thus providing the
basis for a two-arm spectrometer. A vertex pixel detector (inside the
magnetic field) provides
high resolution tracking near the interaction. The baseline design has
seven downstream tracker stations of straw tubes along both arms of
the spectrometer. For identification of electromagnetic final states
and kaons there is electromagnetic calorimetry and a RICH
detector respectively, with a toroidal magnetic detector for muon
identification and measurement. The BTeV detector relies heavily on the
vertex detector for triggering of the \B\ events.

\subsection{Experimental Recap}

 Table~\ref{tab:summary} summarises the experiments discussed previously
in this section. The SLAC and KEK B factories are due to take first data
before the turn of the century.
Though it is important to remember that new accelerators often take 2-3
years to reach their design goals. HERA-B is already partially
instrumented
and is already taking data in situ in order to debug and test some of
their detector components. They have already achieved the multiple 
interaction per bunch crossing that is necessary to meet their design
considerations.  CDF and D0, like the experiments at the $e^+e^-$
machines, are due to take data again before the end of the century.
With their already proven track record in \B\ physics they should be in a
good position to be the first to observe CP violation in the \B\ system.
Beyond that LHC-B and BTEV (in addition ATLAS and CMS) should be in a
good position to further test and even overconstrain the unitarity
triangle of CP violation in the standard model and perhaps any new
physics beyond.

\begin{table}[htb]
\centering
\begin{tabular}{|l|l|c|c|c|}
\hline
& & Collision & Centre of & \\
Experiment & Machine & Type & Mass Energy  & Start\\
\hline
CDF/D0 & Tevatron & $p\bar p$ & 2TeV & 1999 \\
ATLAS/CMS & LHC & $pp$ & 14TeV & 2005\\
& & & & \\
HERA-B & HERA & $pN$ & 40GeV & 1998\\
BaBar & PEP-II & $e^+e^-$ & $\Upsilon(4S)$ & 1999\\
BELLE & KEKB & $e^+e^-$ & $\Upsilon(4S)$ & 1999\\
BTEV & Tevatron & $p\bar p$ & 2TeV  & 2002\\
LHC-B & LHC & $pp$ & 14TeV  & 2005\\
\hline
\end{tabular}
\caption{\it Properties of the experiments}
\label{tab:summary}
\end{table}

\section{CP Reach}
In table~\ref{tab:babar-herab} is a comparison of the statistical
precision that BaBar (and BELLE) and HERA-B claim can be 
achieved in measuring $\sin 2\alpha$ and $\sin
2\beta$ for one year's running at design luminosity. 
The two experiments are comparable in the
$B^0 \rightarrow J/\Psi K_s^0$ channel, but there are other channels that
BaBar can use, because of the clean event environment, to extract the
angle $\alpha.$ For measuring $\sin 2\beta$ at first glance the two
experiments are again comparable, but the figures quoted in the
table are for zero background. HERA-B will have a background in 
$B^0 \rightarrow \pi^+ \pi^-$ channel and the quoted figures need to be
modified by a factor
$\sqrt{1+B/S}$ where $B/S$ is the ratio of background to signal. It is
thought that a value of $B/S < 1$ can be achieved~\cite{abt}.

\begin{table}[hbt]
\centering
\begin{tabular}{|l|c|c|}
\hline
Decay Channel & BaBar & HERA-B \\
\hline
$J/\Psi K_s^0$ evts& & \\
$\Delta \sin 2\beta$& $\pm 0.10$ & $\pm 0.13$\\
All channels& & \\
$\Delta \sin 2\beta$& $\pm 0.06$ & $\pm 0.13$\\
& & \\
$\pi^+ \pi^- $& &\\
$\Delta \sin 2\alpha$& $\pm 0.20$ & $\pm 0.14$ \\
$\rho^{\pm}\pi^{\pm}$& &\\
$\Delta \sin 2\alpha$& $\pm 0.11$ &  \\
All channels& & \\
$\Delta \sin 2\alpha$& $\pm 0.085$ & $\pm 0.14$\\
\hline
\end{tabular}
\caption{\it A comparison of the experimental accuracy of the BaBar and HERA-B
experiments}
\label{tab:babar-herab}
\end{table}

Table~\ref{tab:bhadron} reviews the accuracy that the experiments at
hadron colliders hope to achieve from $10^7$ seconds running. 
Besides the accuracy on $\sin(2\alpha)$ and $\sin(2\beta),$ the measurement
precision and upper limit of $\gamma$ and $x_s$ (comparable
 to $x$ in the $B_d$ system)
are listed, which need the $B_s$ mesons to be
detected and tagged. The figures show that the dedicated experiment at
the LHC, LHC-B, has by far the best reach in measuring the parameters of
CP violation. It is also worth noting that the performance of the
Tevatron general purpose detectors are comparable to result expected from
HERA-B and thus the $e^+e^-$ \B\ factories. (It should be noted though the
clean experimental environment in  $e^+e^-$ allows them to make a more
complete study of rare \B-decays.)

\begin{table}[htb]
\centering
\begin{tabular}{|l|c|c|c|c|c|}
\hline
& CDF/D0 & HERA-B & ATLAS/CMS & LHC-B & BTeV\\
\hline
$\Delta \sin(2\alpha)$& $\sim 0.10$ & $\sim 0.14$ & $0.10/0.07$ & $0.039$ & $0.1$ \\
$\Delta \sin(2\beta)$& $\sim 0.10$ & $\sim 0.13$ & $0.02/0.07$ & $0.023$ & $0.042$\\
$\Delta \gamma$& &  &  & $6-16^{\circ}$ & \\
$x_s$& & $\le 17$ & $\le 34$& $\le 55$ & $\le 30$ \\
\hline
\end{tabular}
\caption{\it A comparison of the experimental accuracy and reach of the
experiments at hadron facilities. All but BTeV data taken from
ref.~\protect\cite{beauty96}. BTeV limits from their
EoI~\protect\cite{BTeVEoI}.}
\label{tab:bhadron} 
\end{table}

\section{Conclusions}
There is a strong program of future experiments planning to study CP
violation in the \B\ system. The first generation experiments start taking
data in the next few years. Second generation experiments are already
being planned to extend the measurement of CP violation. The experiments
will over constrain the unitarity triangle and perhaps indicate new
physics beyond the Standard Model.


\begin{thebibliography}{99}
\parskip=0pt
{\footnotesize
\bibitem{CKM}  M.~Kobayashi and T.~Maskawa, Prog. Theor. Phys. 49 (1973)
652.

\bibitem{Wolfie} L.~Wolfenstein, Phys. Rev. Lett. 51 (1983) 1945.

\bibitem{rosner} J.\ L.~Rosner, `The Cabibbo-Kobayashi-Maskawa Matrix',
in \B\ Decays (World Scientific), ed. S.~Stone.

\bibitem{nir_quinn} Y.~Nir and H.~R.~Quinn, `Theory of CP Violation in
\B\ Decays',  in \B\ Decays (World Scientific), ed. S.~Stone.

\bibitem{dunietz} I.~Dunietz, `CP Violation with Additional \B\ Decays',
in \B\ Decays (World Scientific), ed. S.~Stone.

\bibitem{belletdr} BELLE Collab., M.~T.~Cheng et al.,
Technical Design Report, KEK-Report 95-1.

\bibitem{babartdr} BaBar Collab., D.~Boutigny et al., Technical Design Report,
SLAC-R-95-457.

\bibitem{herabtdr} HERA-B Collab., E.~Hartouni et al.,
Technical Design Report, DESY-PRC-95/01.

\bibitem{cdfup} CDF Collab., R.~Blair et al.,
The CDF II Detector Technical Design Report, FERMILAB-Pub-96/390-E.


\bibitem{d0up} D0 Collab., 
The D0 Upgrade: The Detector and its Physics, Fermilab Pub-96/357-E.

\bibitem{ATLASTDR} ATLAS Collab., W.~W.~Armstrong et al., 
ATLAS Technical Proposal, CERN/LHCC/94-43.

\bibitem{CMSTDR} CMS Collab., G.~L.~Bayatian et al.,
CMS Technical Proposal, CERN/LHCC/94-38.

\bibitem{theseproc} I.~Vichou, `Physics with the ATLAS dtector at LHC',
to appear in these proceedings;  \newline
I.~Efthymiopouos, `Overview of the ATLAS Detector at LHC', to appear in
these proceedings; \newline
G.~Snow, `CMS General Overview and Physics Performance', to appear in
these proceedings; \newline
R.~Ribeiro, `The Tracking System of CMS', to appear in
these proceedings; \newline
D.~Barney, `The CMS Crystal Calorimeter', to appear in
these proceedings.

\bibitem{PYTHIA} T. Sj\"{o}strand, Computer Physics Commun. 39 (1986) 347;
\newline
H.-U. Bengtsson and T. Sj\"{o}strand, Computer Physics Commun. 46 (1987) 43.

\bibitem{LHCBLoI} LHC-B Collab., K.~Kirsebom et al., LHC-B Letter of
Intent, CERN/LHCC 95-5.

\bibitem{BTeVEoI} BTeV Collab.,A.~Santoro et al., BTeV: An Expression
of Interest for a Heavy Quark Program at C0, BTeV-pub-97/2

\bibitem{abt} I.~Abt, Proceedings of Beauty'96, Nucl. Instr. Meth. A384
(1996), 113.

\bibitem{beauty96} Proceedings of Beauty'96, Nucl. Instr. Meth. A384
(1996).
}
\end{thebibliography}
\end{document}